\documentclass[a4paper,fleqn,usenatbib]{mnras}

\usepackage[T1]{fontenc}
\usepackage{ae,aecompl}

\usepackage{graphicx}	
\usepackage{amsmath}	
\usepackage{amssymb}	
\usepackage{txfonts}
\usepackage{enumitem}
\usepackage{natbib}
\usepackage{xcolor}
\usepackage{multirow}
\usepackage{pdflscape}

\title[Atmospheric dispersion]{On-sky measurements of atmospheric dispersion: \\ I. Method validation}

\author[B. Wehbe et al.]{
B. Wehbe,$^{1,2}$\thanks{E-mail: bachar.wehbe@astro.up.pt}
A. Cabral,$^{3,4}$
and G.\'{A}vila
\\
$^{1}$Instituto de Astrof\'{i}sica e Ci\^{e}ncias do Espa\c{c}o, Universidade do Porto, CAUP, Rua das Estrelas, 4150-762 Porto, Portugal\\
$^{2}$Departamento de F\'{i}sica e Astronomia, Faculdade de Ci\^{e}ncias, Universidade do Porto, Rua Campo Alegre, 4169-007 Porto, Portugal\\
$^{3}$Instituto de Astrof\'{i}sica e Ci\^{e}ncias do Espa\c{c}o, Universidade de Lisboa, Campus do Lumiar, Estrada do Pa\c{c}o do Lumiar 22, Edif. D, PT1649-038 Lisboa, Portugal\\
$^{4}$Departamento de F\'{i}sica, Faculdade de Ci\^{e}ncias, Universidade de Lisboa, Campo Grande 1749-016 Lisboa Portugal
}

\date{Accepted XXX. Received YYY; in original form ZZZ}

\pubyear{2020}

\begin{document}
\label{firstpage}
\pagerange{\pageref{firstpage}--\pageref{lastpage}}
\maketitle

\begin{abstract}
	Observations with ground-based telescopes are affected by differential atmospheric dispersion due to the wavelength-dependent index of refraction of the atmosphere. The usage of an Atmospheric Dispersion Corrector (ADC) is fundamental to compensate this effect. Atmospheric dispersion correction residuals above the level of $\sim$ 100 milli-arcseconds (mas) will affect astronomical observations, in particular radial velocity and flux losses. The design of an ADC is based on atmospheric models. To the best of our knowledge, those models have never been tested on-sky. In this paper, we present a new method to measure the atmospheric dispersion on-sky in the optical range. We require an accuracy better than 50 mas that is equal to the difference between atmospheric models. The method is based on the use of cross-dispersion spectrographs to determine the position of the centroid of the spatial profile at each wavelength of each spectral order. The method is validated using cross-dispersed spectroscopic data acquired with the slit spectrograph UVES. We measure an instrumental dispersion of $\rm 47 ~ mas$ in the blue arm, 15 mas, and 23 mas in the two ranges of the red arm. We also measure a 4 \% deviation in the pixel scale from the value cited in UVES manual. The accuracy of the method is $\sim$ 17 mas in the range of 315-665 nm. At this level, we can compare and characterize different atmospheric dispersion models for better future ADC designs.
\end{abstract}

\begin{keywords}
atmospheric effects - instrumentation: spectrographs - methods: data analysis
\end{keywords}



\section{Introduction}
Observations with ground-based telescopes are affected by differential atmospheric dispersion due to the wavelength-dependent index of refraction of the atmosphere. The effects of atmospheric dispersion are well known and discussed by several authors such as \cite{Filippenko1982}, and \citet{Bonsch1998}. An atmospheric dispersion model computes the amount of dispersion taking into consideration the zenithal angle of observation (Z), the pressure (P), the temperature (T), and the relative humidity (RH) at the time of observation. The atmospheric dispersion, in arcseconds in the sky, is given by:
\begin{align}
	\Delta R (\lambda) &= R(\lambda) - R(\lambda_{\rm ref}) \nonumber \\
	\Delta R (\lambda) &\approx 206265 \left[n(\lambda) - n(\lambda_{\rm ref}) \right] \times \tan Z,
	\label{eq:dispersion}
\end{align}
where $R(\lambda)$ is the refraction angle, $n$ is the refractive index, $\lambda_{\rm ref}$ is the reference wavelength, and $Z$ is the zenithal angle of observation. In a slit spectrograph, atmospheric dispersion can create slit losses that will affect the observations as described by \cite{Sanchez2014}. In high-precision astronomical instruments, these effects should be taken into consideration and corrected. \cite{Cuby1998} suggested the use of an atmospheric dispersion corrector (ADC) to counter balance the effects of atmospheric dispersion. In the new generation of fiber-fed high-resolution spectrographs, an imperfect atmospheric dispersion correction operation will introduce a varying slope in the spectral continuum \citep{Pepe2008}. This variation will create flux losses and will affect the final radial veolicty (RV) as shown by \cite{Wehbe2020}. This is due to the fact that slope variation will change the weights of spectral lines used in the computation of RV. When aiming to detect Earth-like planets orbiting Sun-like stars \citep{Fischer2016}, an RV precision of 10 cm s$^{-1}$ is required. To avoid any instrumental errors in high-precision RV meausrements, the ADC residuals should be below the level of 100 milli-arcseconds \cite[mas; see][]{Wehbe2020}. \\
The design of an ADC is based on atmospheric models that, to our knowledge, have never been directly compared to on-sky measurements. The difference between various atmospheric models is severe especially when the observations are carried out in the blue part of the spectrum where the atmospheric dispersion is larger. For example, for a zenithal angle of 60$^{\circ}$, the difference between some of the most used models, Zemax \citep{Zemax} and Filippenko's model \citep{Filippenko1982}, is as large as 50 mas. In fact, \cite{Spano2014} showed that the refractive index, $n(\lambda)$, of different atmospheric models can vary by several 10$^{-8}$ which translates into several hundreds of mas, a value larger than the typical residuals of an ADC. Imperfect atmospheric dispersion correction due to inaccurate models, might result in residuals larger than 100 mas, a value that will affect high-precision observations. With all the improvements in the field of adaptive optics (AO), and the reduction of the size of the image at the entrance of a fiber, the ratio between the amount of dispersion and the size of the image becomes more critical. This will be even more the case on the next generation of extremely large telescopes (ELT) equipped with AO-instruments. Therefore, in order to reduce any atmospheric dispersion effect, it is important to deliver accurate ADC designs with residuals as minimal as possible, and measure on-sky the atmospheric dispersion. To do so, and to be able to reach accurate residuals level that will not introduce RV errors, neither flux losses \citep{Wehbe2020}, we developed a new method to measure atmospheric dispersion on-sky that will help us in characterizing different atmospheric models. The method is explained in section \ref{sec:method}, the sources of errors in section \ref{sec:errors}, and the validation of the method in section \ref{sec:validation}.

\section{Method}
\label{sec:method}
Spectroscopic measurements have the benefit of measuring all the wavelengths of interest at the same time, which eliminates the effect of any change in atmospheric conditions, with the exception of atmospheric dispersion. \cite{Skemer2009} attempted to measure the atmospheric dispersion in the N-band (8260 nm to 11270 nm) using spectroscopy. They used measurements from the spectroscopic mode of the Mid-IR Arrac Camera, generation 4 (MIRAC4) and the Bracewell Infrared Nulling Cryostat \citep[BLINC;][]{Hinz2000} at the 6.5 m MMT. MIRAC4-BLINC (currently unavailable) was equipped with a KRS-5 grism, creating first-order low-resolution (R $\sim$ 100) spectra. Measuring the trace of the grism, by centroiding each wavelength, led them to direct dispersion measures. They concluded that the dominating linear trends in their measurements are in agreement with the infrared atmospheric dispersion model of \cite{Mathar2007}. \\
In this work, we also attempt to measure the dispersion based on spectroscopic observations, albeit using a different approach than the one of \cite{Skemer2009}. The concept is based on the determination of the position of the centroid of the spatial profile at each wavelength of each spectral order in cross-dispersed spectroscopic data. Due to atmospheric dispersion, the image of a point source will be elongated at the entrance of the spectrograph. Using a slit spectrograph, with the slit oriented along the parallactic angle (dispersion direction), results in a dispersed image parallel to the slit. This produces a displacement of the spectrum perpendicularly to the main spectrograph dispersion direction. In a cross-dispersed spectrograph, this results in a displacement of the orders along the cross-dispersion direction. Our idea is to measure the position of the centroid at each point of each wavelength of each spectral order. This will allow us to directly measure the atmospheric dispersion as a function of wavelength. \\
To measure the atmospheric dispersion, a slit spectrograph with no ADC, or with an ADC that can be set to zero dispersion, is needed. The UV-Visual Echelle Spectrograph \citep[UVES;][]{Dekker2000}, installed at the UT2 telescope in Paranal, can fulfill our requirements. We will be able to perform our task by orienting the slit to the parallactic angle direction, and by removing the ADC from the optical path. To be able to fully characterize different atmospheric models, it is important to observe targets at different zenithal angles, hence with different amounts of atmospheric dispersion (Wehbe et al., in preparation). The method we present is not target sensitive. In fact, we can test the method on any target observed with the specific setup described above. Since we are interested mainly in the blue part of the spectra, it is preferable to observe blue stars as they have higher fluxes in the range of interest.

\subsection{Observations}
The observations we analysed were performed between April and May 2019 using UVES (program ID 4103.L-0942(A); PI: B. Wehbe). UVES is equipped with a blue and a red arms. However, due to the optical setup of the spectrograph, the wavelength range is divided into three parts: i) blue arm in the 303-384 nm range, ii) red arm in the 487-567 nm range, and iii) in the 590-665 nm range (the red arm detector is a mosaic of two CCDs). In order to cover a wide range of wavelengths, and to be able to perform an analysis over the optical range of the spectrum, we observed our targets using a dichroic that allowed us to perform the analysis over the wavelength range of 303 nm to 665 nm.  In our analysis we use the blue range starting at 315 nm, as the points below are affected by low atmospheric transmission and low signal-to-noise (S/N) ratios. Table \ref{Table:target} shows the main optical setup used at the time of the observations. Four targets (HD117490 and HD150574 as O-type stars; HD143449 and HD165320 as B-type stars), at different zenithal angles (between 13$^{\circ}$ \& 60$^{\circ}$), were observed. To test the method, we choose the target that was observed at the highest zenithal angle (highest dispersion). HD 117490 was observed with 19 exposures of 20 seconds each (Z between 57$^{\circ}$ \& 60$^{\circ}$). This allowed us to achieve a S/N of 100. This value will allow us to reach an accuracy of $1~\%$ that will lead to a better characterization of different atmospheric models. We tested our method with the four targets and they return similar results. In this paper, for demonstration purposes, we will show the results of only one exposure.

\begin{table}
	\caption{Targets and optical setup at the time of observation. We used dichroic (dic) 1 along with the two cross-dispersers (CD) to cover the wavelength range of interest. We also show the optical setup for the archival data used.}             
	\label{Table:target}      
	\centering
	\begin{tabular}{c c c c}
		Target & Slit direction & ADC & Mode \\
		\hline
		HD 117490 & atmospheric disperion & off & dic\#1 \\
		 & & & CD\#1 \& CD\#3 \\
		 \\
		HD 150574 & atmospheric disperion & off & " \\
		HD 143449 & atmospheric disperion & off & " \\
		HD 165320 & atmospheric disperion & off & " \\
		HD 144470 & atmospheric disperion & off & " \\ 
		\\
		alf Cen & angular separation & off & " \\
		
		\hline
	\end{tabular}	
\end{table}

The analysis presented in the main body of the paper is performed using the blue arm of UVES. The analysis of the red arm data is performed following the same method described in the paper, and we only show the corresponding plots in Appendix \ref{appendix:A}.

\subsection{Data analysis}
\begin{figure}
	\centering
	\includegraphics[width=\hsize]{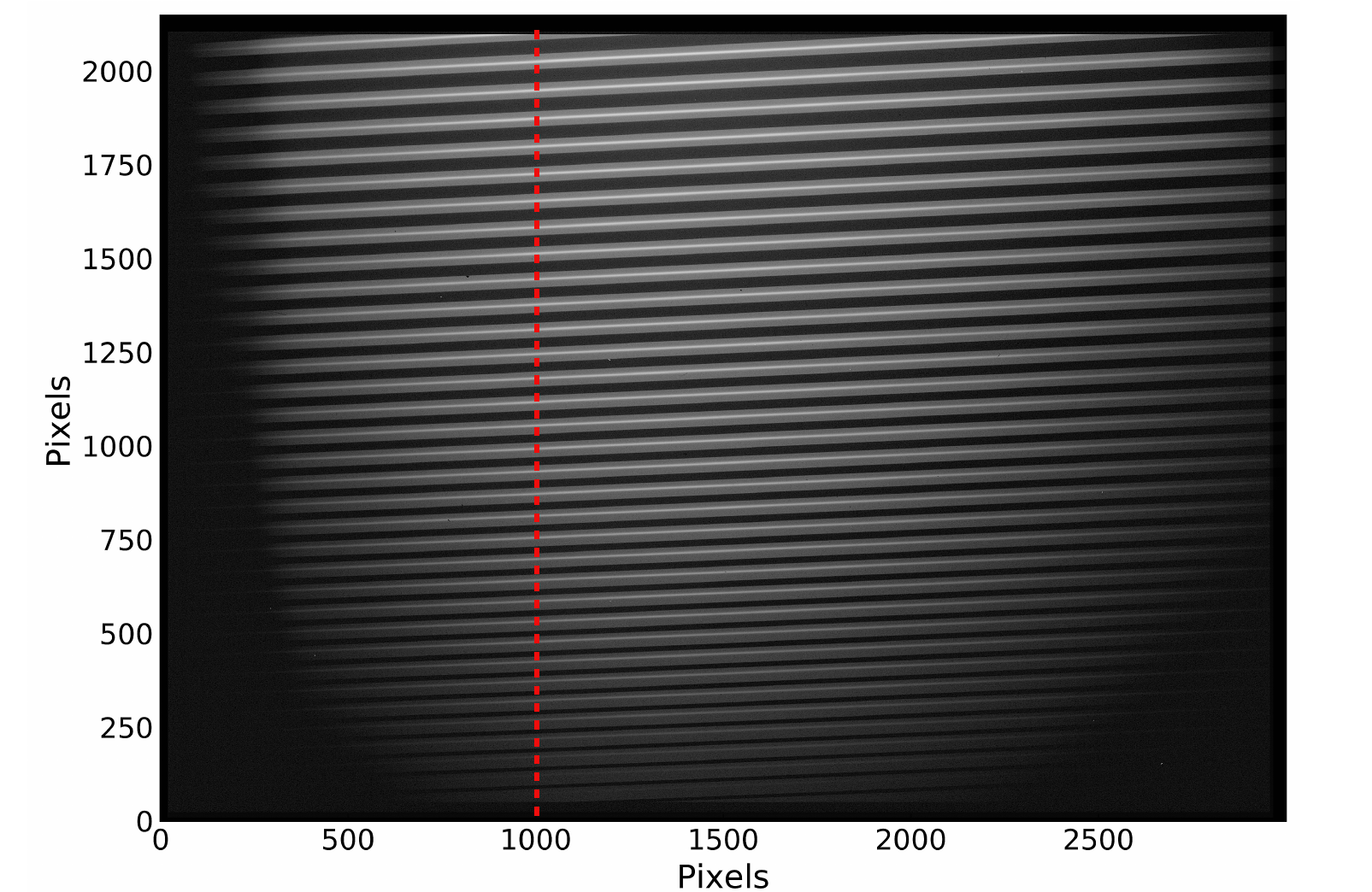}
	\caption{An overlay of the science frame and the flat frame images with 50\% transparency. The thick lines represent the orders of the flat frame; the thin bright lines represent the orders of the science frame. The red dashed line (cut level, used for demonstration purposes in this figure) represents one of the cuts used to compute the centroid.}
	\label{Fig:overlap}
 \end{figure}

\begin{figure*}
	\centering
	\includegraphics[width=\hsize]{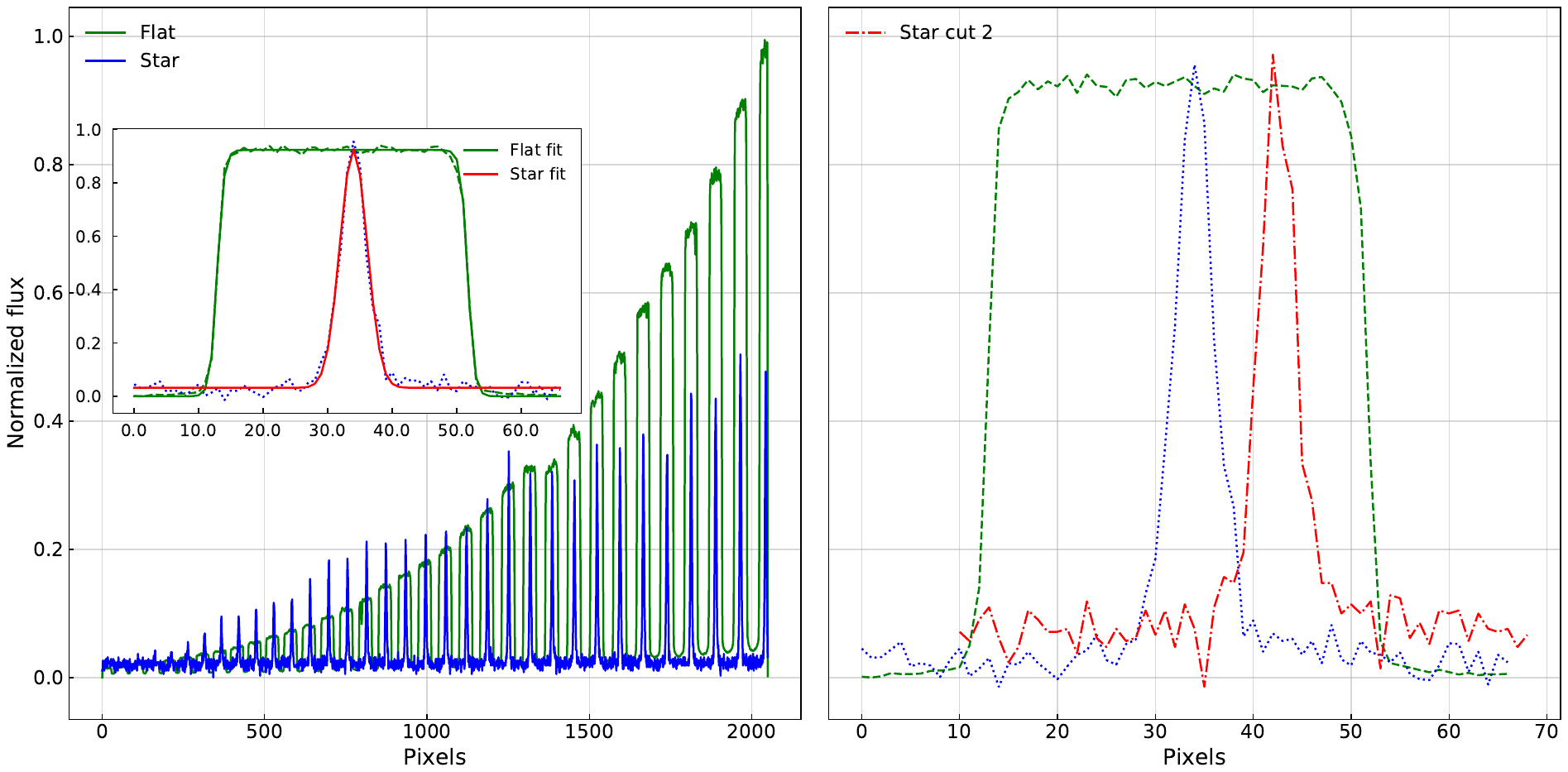}
	\caption{Left: representation of the cut level of Figure \ref{Fig:overlap}. The science frame is overlapping with the flat frame. Each block represents one of the spectral orders along the cut. The inset plot is a zoom in on one of the orders including the fit of the flat and the star. Right: Data from two spectral orders to show the variation of the centroid of the spectral profile.}
	\label{Fig:cut}
\end{figure*}

To measure the position of the centroid accurately, we overlapped the science frame with the flat frame, so we can use it as a reference point (see Figure \ref{Fig:overlap}). The flat frame, obtained by illuminating the full slit with a constant light source, will be used as the zero reference to measure the centroid position variation with the slit. We also made sure that all the different frames were observed with the same instrumental setup (same slit length, dichroic, and filters) in order not to affect the results. The cut level in Figure \ref{Fig:overlap} represents an example of the vertical cuts used to determine the centroids position with respect to the center of the slit, obtained from the flat image, as seen in the left panel of Figure \ref{Fig:cut}.

In order to extract the results from the left panel of Figure \ref{Fig:cut}, we fit the flat data by a Sigmoid function of the form:

\begin{equation}
	S(x) = \dfrac{L}{(1 + e^ {k(x-x_0)}) \times (1 + e^ {-k(x-x_1)})},
	\label{eq:sigmoid}
\end{equation}  

\noindent where $L$ is the amplitude of the data at the centroid, $x_0$ and $x_1$ represent the two edges (left and right) of the function and $k$ is a factor depending on the steepness of the curve. We also fit the star data by a Gaussian function of the form:

\begin{equation}
	G(x) = A \times \left(e^{\dfrac{-(x-x_0)^2}{w}}\right) + c,
	\label{eq:gaussian}
\end{equation}  

\noindent where $A$ is the amplitude of the data, $x_0$ represent the centroid of the gaussian, $w$ the full width at half maximum (FWHM), and $c$ is a constant. The inset in the left panel of Figure \ref{Fig:cut} shows that the Sigmoid and Gaussian functions, are in good agreement with the flat and star data.

\begin{figure}
	\centering
	\includegraphics[width=\hsize]{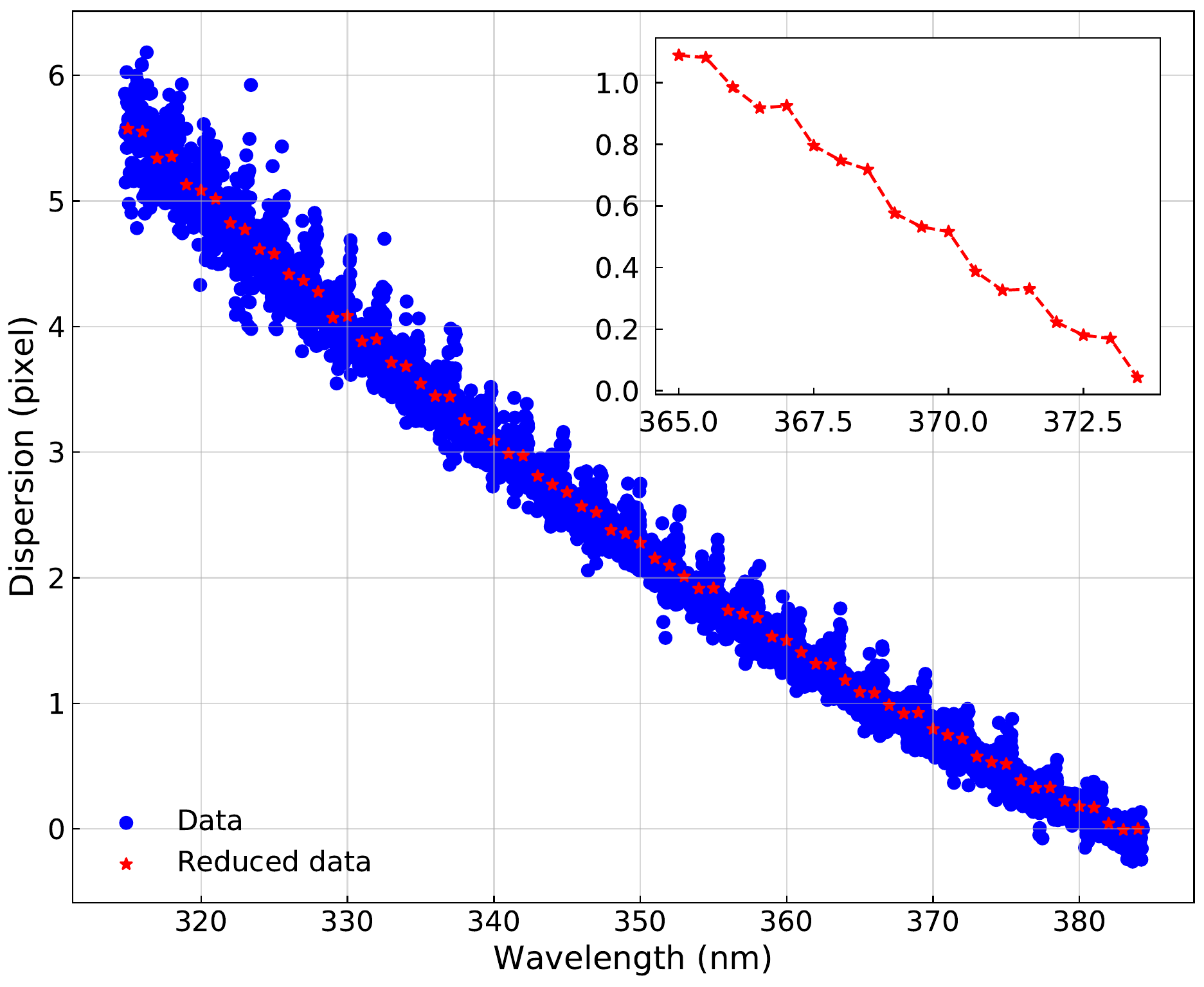}
	\caption{Extracted atmospheric dispersion in pixels using the centroids as a function of wavelength. The red stars represents the reduced version of the data after averaging points for each 1 nm. This plot represents the results for the blue arm only. For the red arm, see Figure \ref{Fig:notreduced_red}.}
	\label{Fig:not-reduced}
\end{figure}

A Gaussian fit is used to extract the position of the centroid of each pixel along the wavelength direction of each order. The Sigmoid fit is used to extract the limits of the slit illuminated in the flat frame. We repeat this step for the same points used in the Gaussian fit. Due to atmospheric dispersion, and to the fact that the wavelengths are dispersed, we detect a variation between the centroid positions as expected (right panel of Figure \ref{Fig:cut}). By repeating the same procedure to all the spectral orders, through all the columns, we will be able to plot the variation of the centroids in terms of pixels. We use the UVES pipeline \citep{uves2019}, to perform the wavelength calibration, to convert the pixels of the $x$-axis to wavelengths. This results in a plot representing the centroid variation as a function of wavelength as shown in Figure \ref{Fig:not-reduced}. This represents a direct measurement of the atmospheric dispersion. Since we are interested in the differential dispersion, we use the center of the slit as a reference point. It is computed from the Sigmoid fit using $((x_0 + x_1)/2)$. Each point in the plot represents the distance between the centroid and the reference point.


To reduce the scatter of the measurements visible on Figure \ref{Fig:not-reduced}, we average the data every 1 nm. The results are shown as red stars in Figure \ref{Fig:not-reduced}. The same figure also shows a periodicity that is most likely caused by the order separation. This periodicity is also imprinted on the averaged data as shown in the zoomed in plot. The order separation in the blue arm is smaller than the one of the red arm orders. This is represented in longer periodicities in the plots of Figure \ref{Fig:notreduced_red}. Similarly, the red arm results are shown in Figure \ref{Fig:notreduced_red}. In the next section, we explore the various sources of errors on the meausrements.

\section{Sources of errors in sky-dispersion measurements}
\label{sec:errors}
In this work, we present a method to measure the on-sky atmospheric dispersion. This will allow us to characterize different atmospheric dispersion models with an accuracy better than 50 mas. Measurement uncertainties are associated with errors that define the confidence interval of the results. In this section, we describe the sources of errors that contribute to the final uncertainty, in particular the ones related to the pixel scale and the instrumental dispersion. In Section \ref{sec:uncertainties}, we present a treatment of the random errors that additionally exist in the data. We also present the overall uncertainty budget of the measurements.

\subsection{Pixel scale}
\label{subsec:pix}
The pixel scale is the angular distance between two objects on the sky that fall one pixel apart on the detector (CCD), having units of mas/pixel. On the CCD of a slit-fed spectrograph, we can define two pixel scales: i) along the main dispersion direction (varying along the order) which represents the wavelength dispersion (x-axis); and ii) along the slit and the cross-dispersion (y-axis). Due to optical distortions, both scales may vary slightly across the detector. We are interested in the y-axis pixel scale as the atmospheric dispersion is along this axis when the slit is oriented along the parallactic angle. For UVES, taking into consideration the instrumental setup and the CCD used during our observations, the pixel scale listed in the manual \citep{uves2018}, $\rm PS_{m}$, is 250 mas/pixel and $\rm 180 ~ mas/pixel$ for the blue and red arms, respectively. The $\rm PS_{m}$, specifically the one along the slit, is an average value and most likely measured during the commissioning of the instrument, 20 years ago, with an accuracy that is not enough for our requirements. A deviation of the order of 3 \% can cause a variation of 50 mas, the accuracy we are trying to achieve. It is crucial for our work to be able to measure accurately this pixel scale, in order to convert the dispersion in pixels (Figure \ref{Fig:not-reduced}) into mas on the sky. \\ We describe and compare two different methods used to measure and analyze the pixel scale in the following subsections: one based on the separation of a binary system, while the other is based on the slit length. When dealing with the red part of the spectrum (see Appendix \ref{appendix:A}), since red data of the binary system was not available, we will have to use slit length method.

\subsubsection{Binary system method}
\begin{figure}
	\centering
	\includegraphics[width=\hsize]{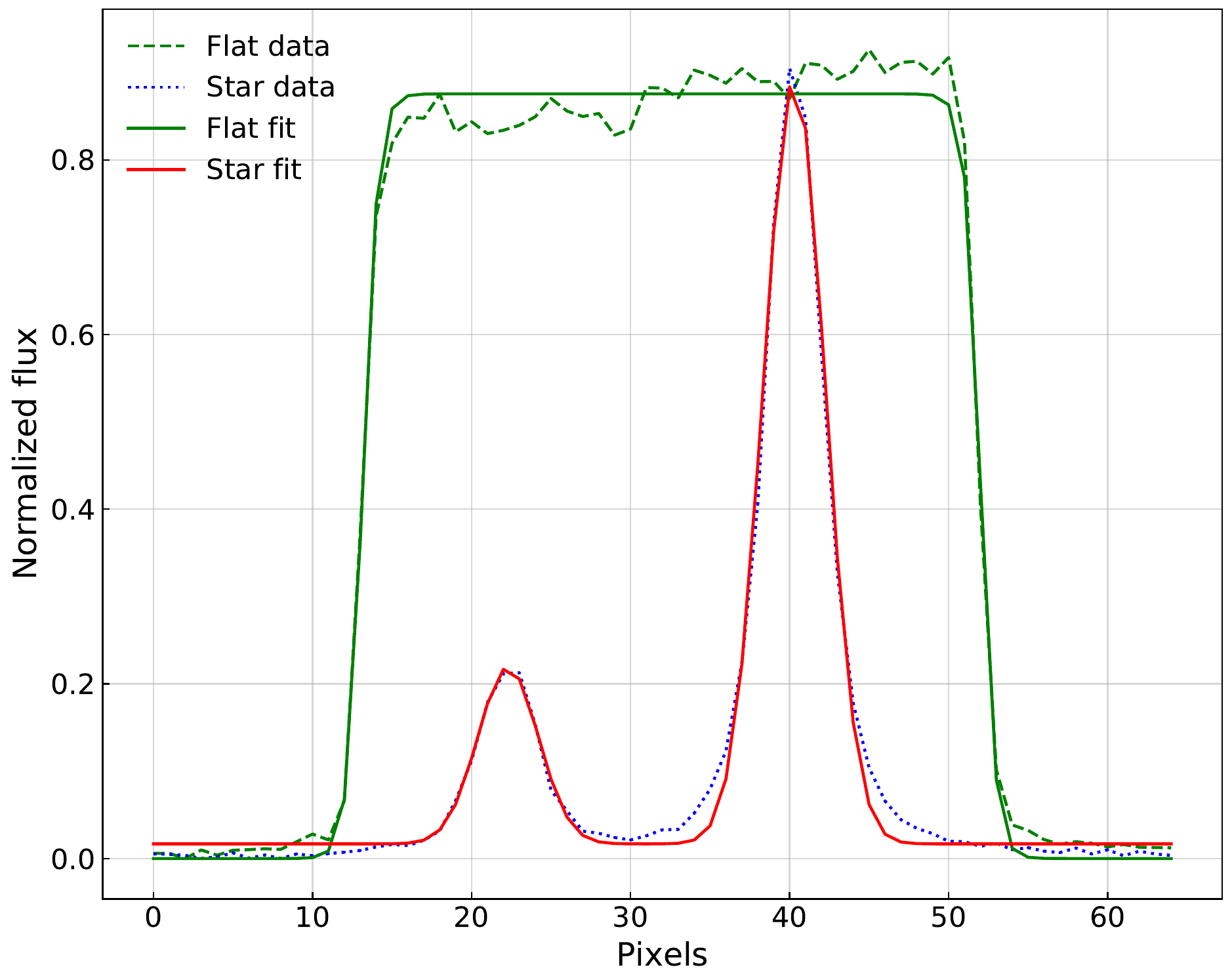}
	\caption{Zoom in on one of the orders of the binary system showing the two peaks separation.}
	\label{Fig:mos}
\end{figure}

One of the methods to determine the pixel scale with an accuracy better than 3 \% is to observe a binary system with a very well measured angular separation between the two targets. To be able to use this separation to measure the pixel scale, it is required that the slit is oriented along the angular separation between the two targets. Archival data of Alpha Centauri observed with UVES (program ID 091.C-0838(A)) at Z = 48$^{\circ}$, using the same instrumental setup as ours, and with the slit oriented along the angular separation, are present in the ESO archive (see Table \ref{Table:target}). We use the same method described in section \ref{sec:method} to determine the separation between the two peaks in pixels for every point on the CCD. Knowing the angular separation of the two targets in mas, $\rho$ (see Appendix \ref{appendix:B}), we can convert the pixel separation into pixel scale. In Figure \ref{Fig:mos}, we show the data and the fit of one order of a cut. Using a two-peak Gaussian function similar to equation \ref{eq:gaussian}, we can easily determine the separation in pixels which is varying between 17.92 and 18.15 pixels in the blue range (315 nm to 384 nm). The angular separation, on-sky, between the binary, is not expected to vary within the observation. In the event of atmospheric disperison, the spectra of the two stars will be affected in the same manner. Therefore, their angular separation will remain constant, and wavelength-independent. Hence, this will not affect our results. The angular separation is computed following the procedure explained in \cite{Duffett2017}, and presented in Appendix \ref{appendix:B}. The calculation of a binary system orbit proceeds in much the same way as that of a planetary orbit by solving Kepler's equation. To do so, we use the orbital solutions (period of revolution, eccentricity of the orbit, semi-major axis of the orbit, inclination...) of Alpha Centauri from \cite{Pourbaix2016}. Taking into account the time of observations of the target, we computed an angular separation of $\rho = 4680~\pm~1$ mas. This value indicates a pixel scale variation in the blue range between 258 mas/pixel and 261 mas/pixel. We note that this procedure is independent of the selected binary target. This means that any binary system, with the slit oriented along their separation will give the same results in terms of pixel scale. 

\subsubsection{Slit length method}
\begin{figure}
	\centering
	\includegraphics[width=\hsize]{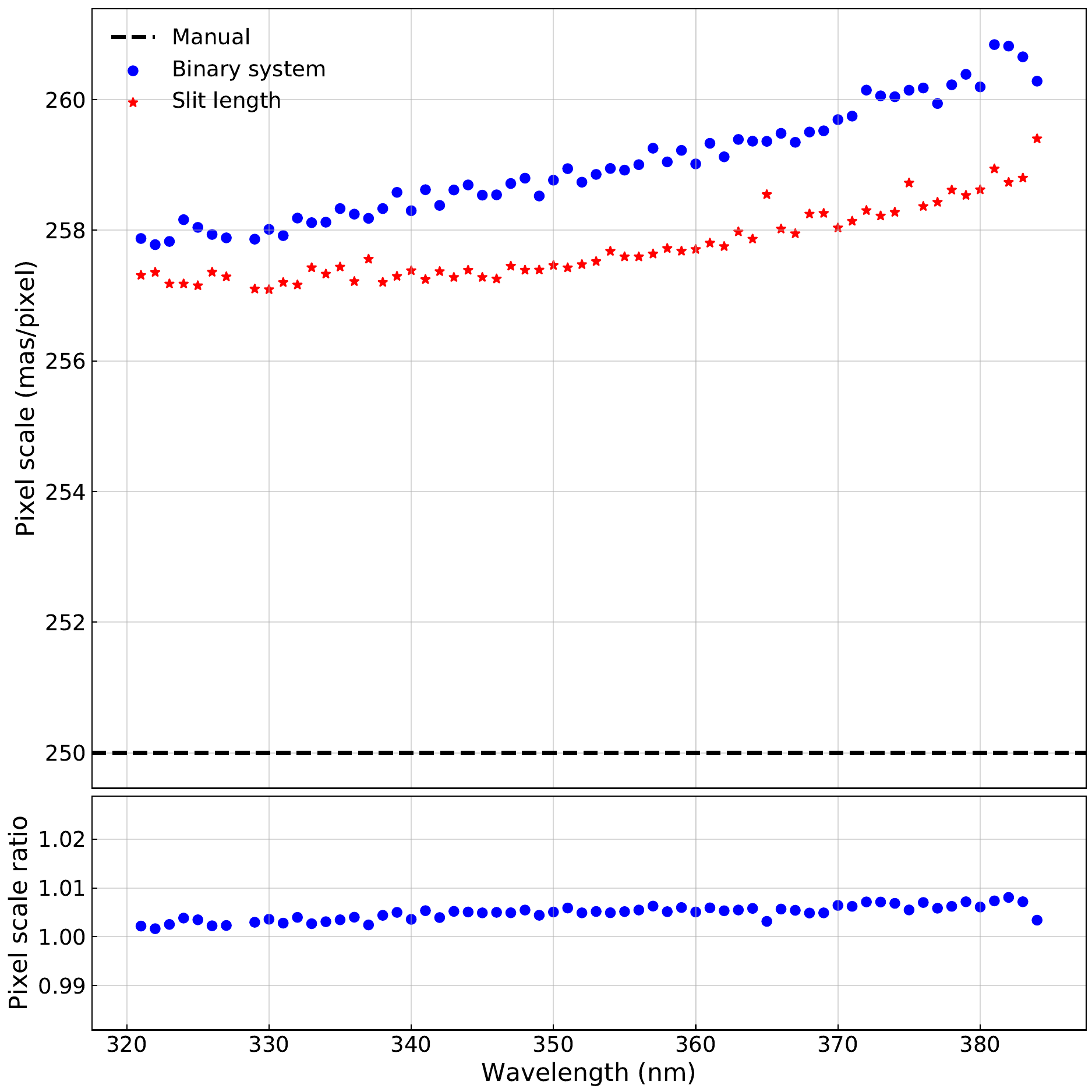}
	\caption{Top: Pixel scale computed for the two methods described above. The dashed line corresponds to the pixel scale value cited in the manual of UVES \citep{uves2018}; bottom: Ratio of the computed pixel scale for the two methods described above.}
	\label{Fig:ps}
\end{figure}

In the slit length method, instead of measuring the separation of a binary system, we rely on the measurements of the slit size using the flat field frames. The slit length in UVES, in the blue arm, is 10000 mas, and known to an accuracy better than 100 mas (private communication with Luca Sbordone, UVES instrument scientist). Knowing the slit length in mas from the UVES manual, and in pixels using the Sigmoid function (equation \ref{eq:sigmoid}) applied to the flat field as shown in the left panel of Figure \ref{Fig:cut}, we will be able to also measure the pixel scale for every point on the CCD and extract it as a function of wavelength. \\
In the top panel of Figure \ref{Fig:ps}, we show the results of the pixel scale for both methods in the blue channel. We plot also the $\rm PS_{m}$ (dashed line) for comparison. The slit length method returns a pixel scale variation in the blue range between 257 mas/pixel and $259 ~\rm mas/pixel$. In the bottom panel of Figure \ref{Fig:ps}, we show the ratio of the computed pixel scale between the two aforementioned methods. This indicates that the two methods are consistent within $0.5~\%$ that is used to calibrate the slit length method in the red range. In the rest of this work, we will use the pixel scale computed using the binary system method in the blue range, and the one computed using the calibrated slit length method in the red range. Note that the pixel scale is wavelength dependent.

%

\begin{figure}
	\centering
	\includegraphics[width=\hsize]{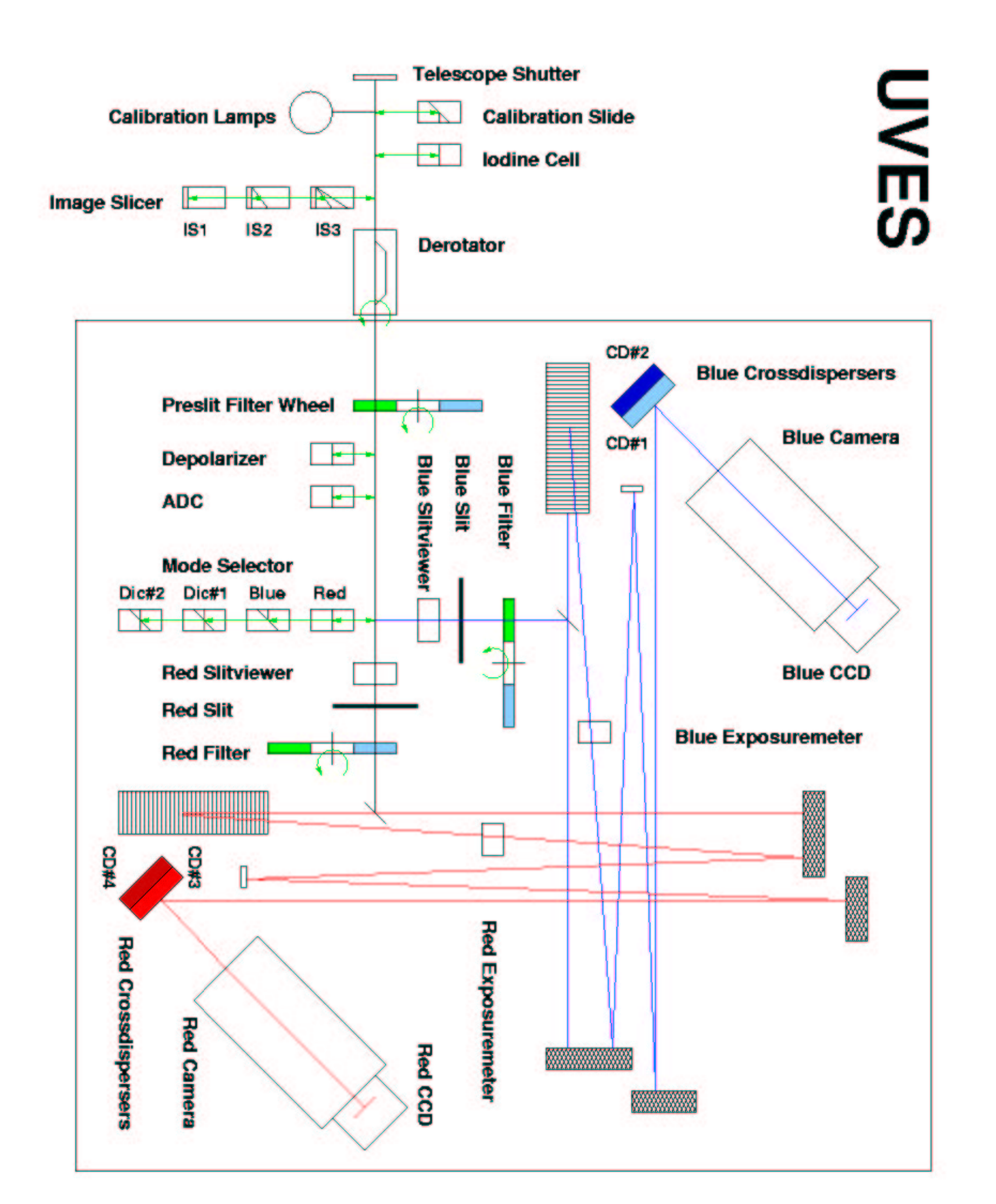}
	\caption{Schematic overview of the UVES spectrograph \citep[adapted from][]{uves2018}.}
	\label{Fig:uves}
\end{figure}

\subsection{Instrumental dispersion}
\label{subsec:inst}
\begin{figure}
	\centering
	\includegraphics[width=\hsize]{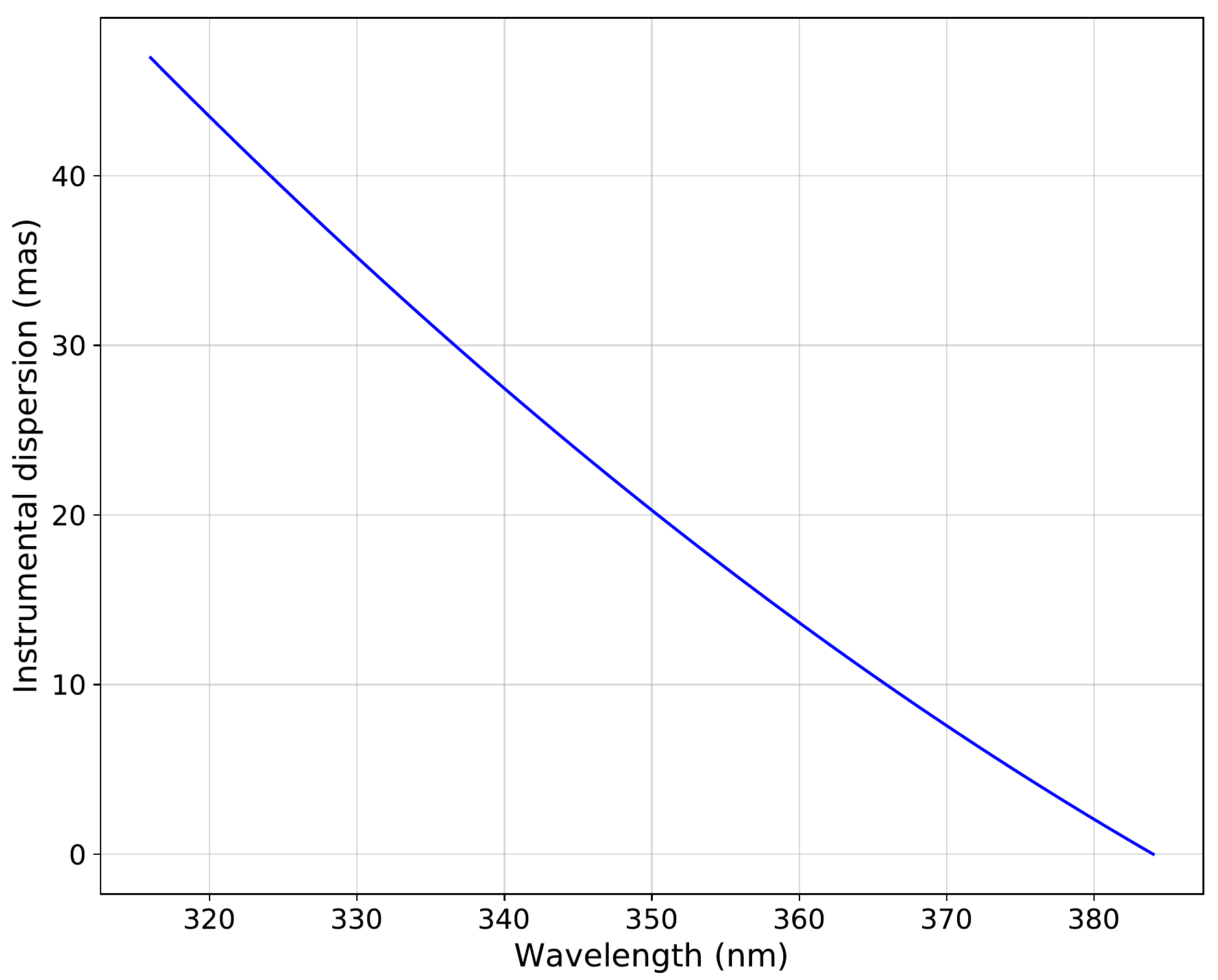}
	\caption{Fit of the instrumental dispersion of UVES using our setup, in the blue arm. The results obtained for the red arm are shown in Figure \ref{Fig:instrumental_red}}.
	\label{Fig:instrumental_blue}
\end{figure}

Optical components, and in particular prisms, can introduce instrumental dispersion that will affect the measurements. In order to accurately test our method with an atmospheric model, it is important to check if any instrumental dispersion is altering the atmospheric one. To do so, we used our technique to extract the centroid variation, but with targets observed at an airmass of 1, where the atmospheric dispersion is null. Based on data obtained from the ESO Science Archive Facility (program ID 194.C-0833(C)), we were able to find data for the target HD144470 (B-type star) observed through UVES, with the same instrumental setup as our observations, at an airmass of 1 (see Table \ref{Table:target}). Following the same method to extract the dispersion in pixels, and using the pixel scale we have computed, we were able to detect a non-zero instrumental dispersion. Given the schematic of UVES (Figure \ref{Fig:uves}), it is most likely that an instrumental dispersion will be introduced due to the derotator used in our observations. The derotator of UVES is an Abbe-Koenig type silica prism that is placed in the diverging beam of the telescope and provides compensation for the field rotation. In fact, both calibration and science frames were observed using the same instrumental setup (same optical components in the light path). Since we were interested in aligning the slit along the atmospheric dispersion direction in our science frames, the derotator was set to the corresponding mode "ELEV" as stated in the manual. In this mode, the derotator will rotate in a way to keep the slit aligned to the atmospheric dispersion direction. In addition, in the flat frames, the derotator was set to another mode, "SKY", where the slit is set to a fixed position (different than the one of the science frames). This difference in the derotator position, between the science and flat frames, introduces an extra dispersion in the direction of the atmospheric one in our spectra. We computed a peak-to-valley (PTV) instrumental dispersion of 47.2 $\pm$ 8.3 mas, 15.2 $\pm$ 6.1 mas, and 23.1 $\pm$ 5.5 mas for the blue and the two red ranges, respectively. The positions of the derotator between the different science datasets were approximately constant, introducing a variation of the instrumental dispersion below 3 mas. The uncertainty on the instrumental dispersion represents the standard deviation of the mean of all the data used, as well as the derotator positions variation. Figures \ref{Fig:instrumental_blue} and \ref{Fig:instrumental_red} show the instrumental dispersion in the blue and red ranges, respectively.

\subsection{Uncertainties}
\label{sec:uncertainties}
In order to test different atmospheric dispersion models with an accuracy better than 50 mas, it is fundamental to evaluate the accuracy of the method used. For this evaluation, we take into account all the sources of errors (systematic and random). To do so, and for simplification purposes, we show the case of one wavelength $\rm (315 ~ nm)$ of one exposure (the one used in this paper, see Tables \ref{Table:target} - \ref{Table:atm}) as an example to compute the uncertainties related to the sources of errors detailed in subsections \ref{subsec:pix} and \ref{subsec:inst}. In this subsection, we describe the main sources of uncertainties, and provide an example in Table \ref{Table:uncertainties}, including the total uncertainty that should be taken into consideration in our work. Note that in Table \ref{Table:uncertainties} we provide the error budget of the pixel scale using the two methods mentioned in subsection \ref{subsec:pix}, while we use only the results of the binary system separation method. The final uncertainty computed represent the error bars shown in Figure \ref{Fig:blue_all}. It can be seen that the uncertainties are wavelength dependent. Therefore, the same procedure explained below and shown in Table \ref{Table:uncertainties}, is applied at all the wavelengths. These results represent the 1 $\sigma$ (68.3 \%) confidence level. The uncertainties are computed as mentioned in the "\textit{Guide to the expression of uncertainty in measurements}" \citep{Gum1995}, and follow equation \ref{eq:uncertainty}.

\begin{equation*}
	y = f(x_{\rm 1},x_{\rm 2},...,x_{\rm N})
\end{equation*}

\begin{equation}
	u_{\rm c}(y) = \sqrt{\sum_{\rm i=1}^{N} u^2(x_{\rm i})\left(\frac{\partial f}{\partial x_{\rm i}}\right)^2},
	\label{eq:uncertainty}	
\end{equation}
where $y$ is the measured component, $x_{\rm i}$ for i in the range [1,N] are its sub-components, $u(x_{\rm i})$ is the standard uncertainty of each sub-component, and $u(c)$ is the combined uncertainty of each component. The different uncertainties components are divided into three main parts:
\begin{itemize}
	\item Pixel scale calibration related to the binary and slit method used. The two methods have similar sub-components:
	\begin{enumerate}
		\item accuracy of the fits used (Gaussian \& Sigmoid): we generate Gaussian (and Sigmoid) data with similar noise compared to the on-sky data and fit them using the corresponding functions. The variation of the centroid results represent the accuracy of the fits. It is equal to 0.05 pixel in our example.
		\item accuracy of the binary separation: this value is extracted from \cite{Pourbaix2016} and is equal to 1 mas (0.02 \% of the total angular separation, $\rho$ = 4680 mas).
		\item accuracy of the slit length: this value is extracted from \citet{uves2018} and is equal to 50 mas (0.5 \% of the slit length, 10000 mas). This accuracy, as well as the one of the binary separation, are considered in our measurements as rectangular distributions, by excess, where we account for the $\sqrt{3}$ factor to convert them into normal distributions.
		\item standard deviation: as stated above, each point is the mean of all the data of 1-nm bins. In the example we show, we had 49 data points to average. We consider the standard deviation of the mean as another source of uncertainty. It is equal to 0.02 pixel in our example.
	\end{enumerate}
	Using equation \ref{eq:uncertainty}, we can derive the combined uncertainty on the pixel scale. It is equal to 0.87 mas using equation \ref{eq:up}.

	\begin{equation}
		u_{\rm pscale} = \sqrt{\left(u_{1}^2\right)\left(\dfrac{1}{D_{\rm B_p}}\right)^2 + \left(u_{2}^2 + u_{3}^2\right)\left(\dfrac{-D_{\rm B_a}}{D_{\rm B_p}^2}\right)^2},
		\label{eq:up}
	\end{equation}
	where $u_1$ is the uncertainty on the angular separation in mas, $u_2$ and $u_3$ are the uncertainties of the fits and the standard deviation in pixels, respectively, $D_{\rm B_p}$ is the distance of the binary system in pixels (18 pixels in our example), and $D_{\rm B_a}$ is the distance of the binary system in mas (4680 mas in our example).

	\item Instrumental dispersion calibration: The uncertainty of this component is the standard deviation of the measurements (at airmass 1), and is considered constant over the range. It is equal to 8.3 mas.
	
	\item Atmospheric dispersion measurement: This component has similar sub-components as the ones of the accuracy of the fits as well as the standard deviation. In addition, we take into consideration the total uncertainty of the pixel scale (Equation \ref{eq:up}). The combined uncertainty on the atmospheric dispersion measurement is also derived from equation \ref{eq:uncertainty}. It is equal to 16.18 mas using equation \ref{eq:ua}.
	
	\begin{equation}
		u_{\rm Disp} = \sqrt{\left (u_{\rm pscale}^2 \right)\left (D_{\rm AD_p} \right)^2 + \left (u_{2}^2 + u_{3}^2 \right)\left (PS \right)^2},
		\label{eq:ua}
	\end{equation}
	where $D_{\rm AD_p}$ is the measured atmospheric dispersion in pixels ($\rm 5 ~ pixels$ in our example), and $PS$ is the computed pixel scale in mas.
	
\end{itemize}

The final uncertainty of the method listed, taking into consideration the components described above, is computed using equation \ref{eq:ut}.

\begin{equation}
	u_{\rm final} = \sqrt{\left (u_{\rm Disp} \right)^2 + \left (u_{\rm instrumental} \right)^2}
	\label{eq:ut}
\end{equation}

The accuracy of the method is as follows: $\pm$ 18 mas in the blue range, $\pm$ 17 mas in the red range. These results are for a 1 $\sigma$ interval. In the blue channel, this accuracy corresponds to approximately $1~\%$ of the total expected atmospheric dispersion at Z = 60$^{\circ}$ ($\rm 1500 ~ mas$), a value lower than our requirements (3 \%).

\section{Method validation}
\label{sec:validation}
After measuring the atmospheric dispersion in pixels, and quantifying the non-negligible instrumental dispersion, we are now able to convert the measured dispersion from pixels to mas in the sky using the pixel scale. The measured atmospheric dispersion should be validated using atmospheric models with inputs as the same atmospheric parameters at the time of observation (see Table \ref{Table:atm}). We use the Filippenko's model \citep{Filippenko1982} as a test model since it is commonly used to compute the atmospheric dispersion.

\begin{table}
	\caption{Atmospheric parameters at the time of observation.}             
	\label{Table:atm}      
	\centering
	\begin{tabular}{c c c c}
		Temperature ($^{\circ}$C) & Pressure (mbar) & RH (\%) & Zenithal angle ($^{\circ}$) \\
		\hline
		12.31 & 742.58 & 4.5 & 57.5 \\
		\hline
	\end{tabular}
\end{table}

\begin{figure}
	\centering
	\includegraphics[width=\hsize]{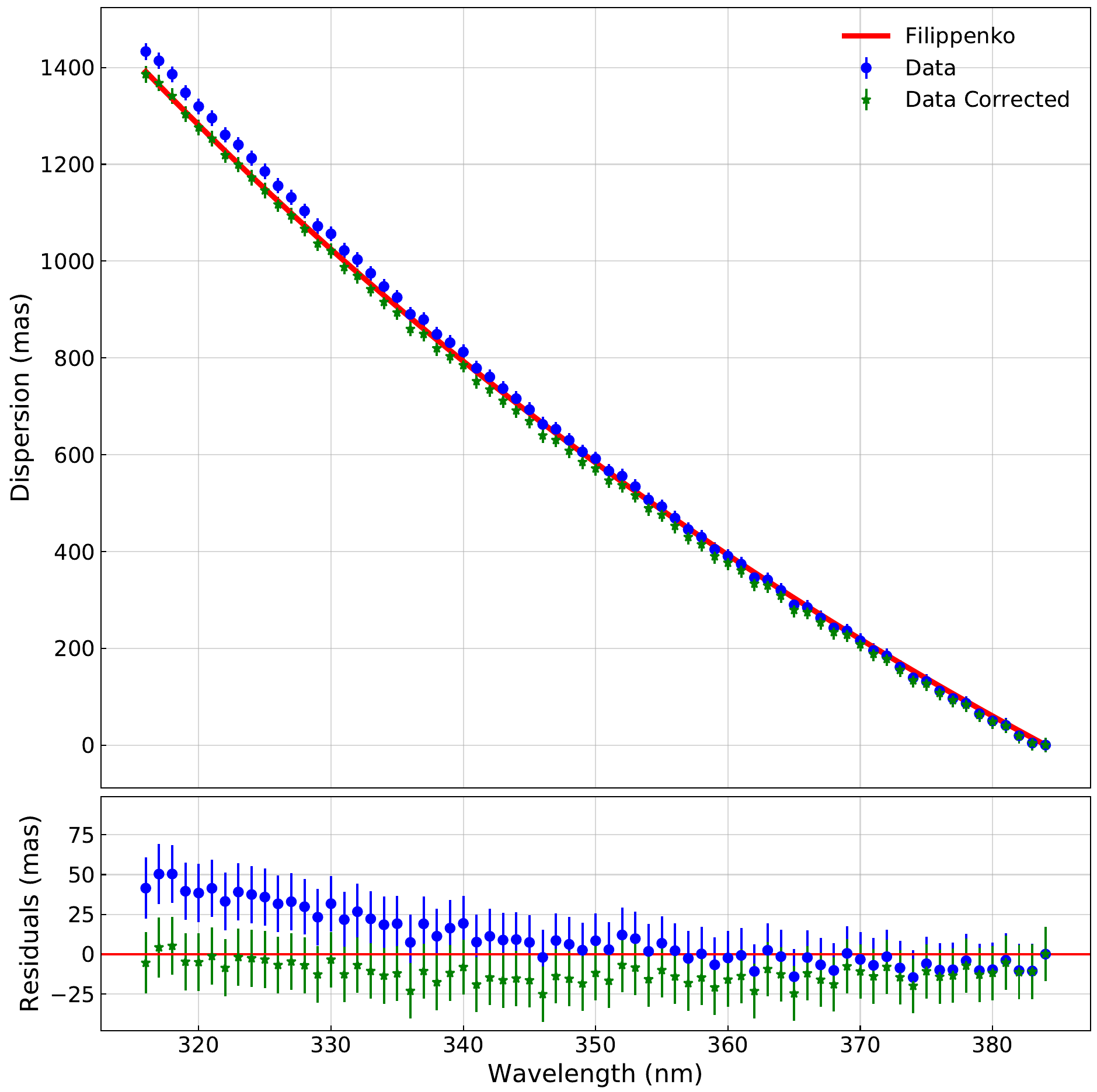}
	\caption{Top: measured atmospheric dispersion, before and after correction by the instrumental dispersion in the blue range. We also show the expected dispersion using Filippenko's model; bottom: atmospheric dispersion residuals, before and after correction by the instrumental dispersion. The results obtained for the red arm are shown in Figures \ref{Fig:red_all1} \& \ref{Fig:red_all2}.}
	\label{Fig:blue_all}
\end{figure}

The top panel of Figure \ref{Fig:blue_all} shows the measured atmospheric dispersion (in mas) in the blue range, before and after subtracting the instrumental dispersion. We also include the uncertainties computed in Table \ref{Table:uncertainties} as error bars. In the same panel, we plot the expected atmospheric dispersion from the Filippenko's model. The bottom panel of Figure \ref{Fig:blue_all} shows the residuals (data - model) for the two cases, before and after instrumental dispersion subtraction. It is clear that at shorter wavelengths, there is an enhancement in the results as the model fits better the corrected data. At longer wavelengths, the two results are consistent within error bars. When looking at the residuals in the bottom panel of Figure \ref{Fig:blue_all}, we can clearly see that the residuals of the corrected data are smaller, and more stable, which is important when validating atmospheric models. In fact, after correction, the PTV of the residuals dropped from 65 mas to 30 mas. In the upper part of the red arm (see Figure \ref{Fig:red_all1}), there was no improvement in the results after correction but they remain consistent within the error bars. In the upper part of the red arm (see Figure \ref{Fig:red_all2}) we have similar results as the blue arm. After the correction, the residuals dropped from 24 mas to 9 mas.

\section{Conclusions}

We present a method to measure on-sky the atmospheric dispersion. The method is based on the determination of the position of the centroid of the spatial profile at each wavelength. Although it is difficult to build an instrument dedicated to do so, and even that UVES showed an instrumental dispersion, the method shows potential in characterizing atmospheric dispersion models. The goal behind testing different models is to reduce any atmospheric dispersion residuals due to differences between the models. As mentioned before, different atmospheric models can lead to a difference of 50 mas, a value bigger than the ADC residuals required for current state of the art spectrographs. From this paper, we find the following:

\begin{enumerate}
	
	\item The method is validated in the optical range (315-665 nm) where the dispersion is more severe. It can also be used for any spectral range and with any slit spectrograph.
	\item We detect an instrumental dispersion in UVES, due to the derotator when at ''ELEV" mode. It is equal to 47 mas in the blue arm, 15 mas and 23 mas in the lower and upper ranges of the red arm, respectively. This extra dispersion should be corrected when measuring on-sky atmospheric dispersion.
	\item The method is tested using Filippenko's model. Using this model, we measure residuals at the level of 30 mas in the blue range as well as the lower red range. In the upper red range, we measure residuals at the level of 9 mas.
	\item The total computed uncertainties of the method, for a 1 $\sigma$ interval, are around 17 mas for the blue and the red ranges. This accuracy will allow us to better characterize different atmospheric models in order to return better ADC designs in the future (Wehbe et al., in preparation).
	
\end{enumerate}

\begin{landscape}

	\begin{table}
		\caption{Uncertainties budget showing in details what each source of error is introducing. The contribution of each sub-source is computed using equation \ref{eq:uncertainty} and show the results at the wavelength 315 nm of one exposure. The total uncertainty of each source is computed using the root sum squared of its contributions. The last column represnts the references used in these computations: 1: \protect\cite{Pourbaix2016}; 2: study with generated data; 3: experimental standard deviation of the mean value; 4: \protect\cite{uves2018}.}             
		\label{Table:uncertainties}      
		\centering                          
		\begin{tabular}{c c c c c c}        
		\hline                
		Source of uncertainty & Sub-source & Uncertainty & Probability Distribution & Contribution & Reference \\
		\hline
		\multirow{3}{*}{\textbf{Pixel scale using binary separation}} & angular separation & 1 mas & rectangular & 0.032 mas & 1 \\
		& accuracy of the fit & 0.05 pixel & normal & 0.82 mas & 2 \\
		& standard deviation & 0.02 pixel & normal & 0.28 mas & 3 \\
		& \multicolumn{3}{c}{\textbf{total ps using binary separation uncertainty}} & \textbf{0.87 mas} &  \\
		\hline
		\multicolumn{6}{c}{} \\
		\multirow{3}{*}{\textbf{Pixel scale using slit length distance}} & slit length & 50 mas & rectangular & 0.75 mas & 4 \\
		& accuracy of the fit & 0.05 pixel & normal & 0.38 mas & 2 \\
		& standard deviation & 0.02 pixel & normal & 0.13 mas & 3 \\
		& \multicolumn{3}{c}{\textbf{total ps using slit length distance uncertainty}} & \textbf{0.85 mas} &   \\
		\hline
		\multicolumn{6}{c}{} \\
		\multicolumn{2}{l}{\textbf{Instrumental dispersion}} & 8.30 mas & normal & \textbf{8.30 mas} & 3 \\
		\hline
		\multicolumn{6}{c}{} \\
		\multirow{3}{*}{\textbf{Atmospheric dispersion}} & pixel scale & 0.87 mas & normal & 4.33 mas & computed \\
		& accuracy of the fit & 0.05 pixel & normal & 14.70 mas & 2 \\
		& standard deviation & 0.02 pixel & normal & 5.20 mas & 3 \\
		& \multicolumn{3}{c}{\textbf{total atmospheric dispersion uncertainty}} & \textbf{16.18 mas} &  \\
		\hline
		\multicolumn{6}{c}{\textbf{accuracy of the method: 18 mas}} \\
		\hline                       	
		\end{tabular}
	\end{table}
\end{landscape}

\section*{Acknowledgements}
The first author is supported by a Funda\c{c}\~{a}o para a Ci\^{e}ncia e Tecnologia (FCT) fellowship (PD/BD/135225/2017), under the FCT PD Program PhD::SPACE (PD/00040/2012). This work was supported by FCT/MCTES through national funds and by FEDER - Fundo Europeu de Desenvolvimento Regional through COMPETE2020 - Programa Operacional Competitividade e Internacionaliza\c{c}\~{a}o by these grants: UID/FIS/04434/2019; PTDC/FIS-AST/32113/2017 \& POCI-01-0145-FEDER-032113.

\section*{Data availability}
The data underlying this article are based on observations made with ESO Telescopes at the Paranal Observatory under program ID 4103.L-0942(A) (with the UVES spectrograph at the ESO VLT UT2 telescope) and available in ESO Science Archive Facility at http://archive.eso.org/, and can be accessed with the program ID.




\bibliographystyle{mnras}
\bibliography{wehbereferences} 




\appendix
\section{Red arm}
\label{appendix:A}
The work presented above shows the plots of the blue arm only. We used the same procedures and techniques to work on the red arm as well. We show here only the plots of the same target used in the paper, HD 117490. The CCD detector in the red arm consists of a mosaic of two chips, separated by a gap. This is why, each plot is actually a set of 2, showing the results of the upper and lower chip.

\begin{figure}
	\centering
	\includegraphics[width=\hsize]{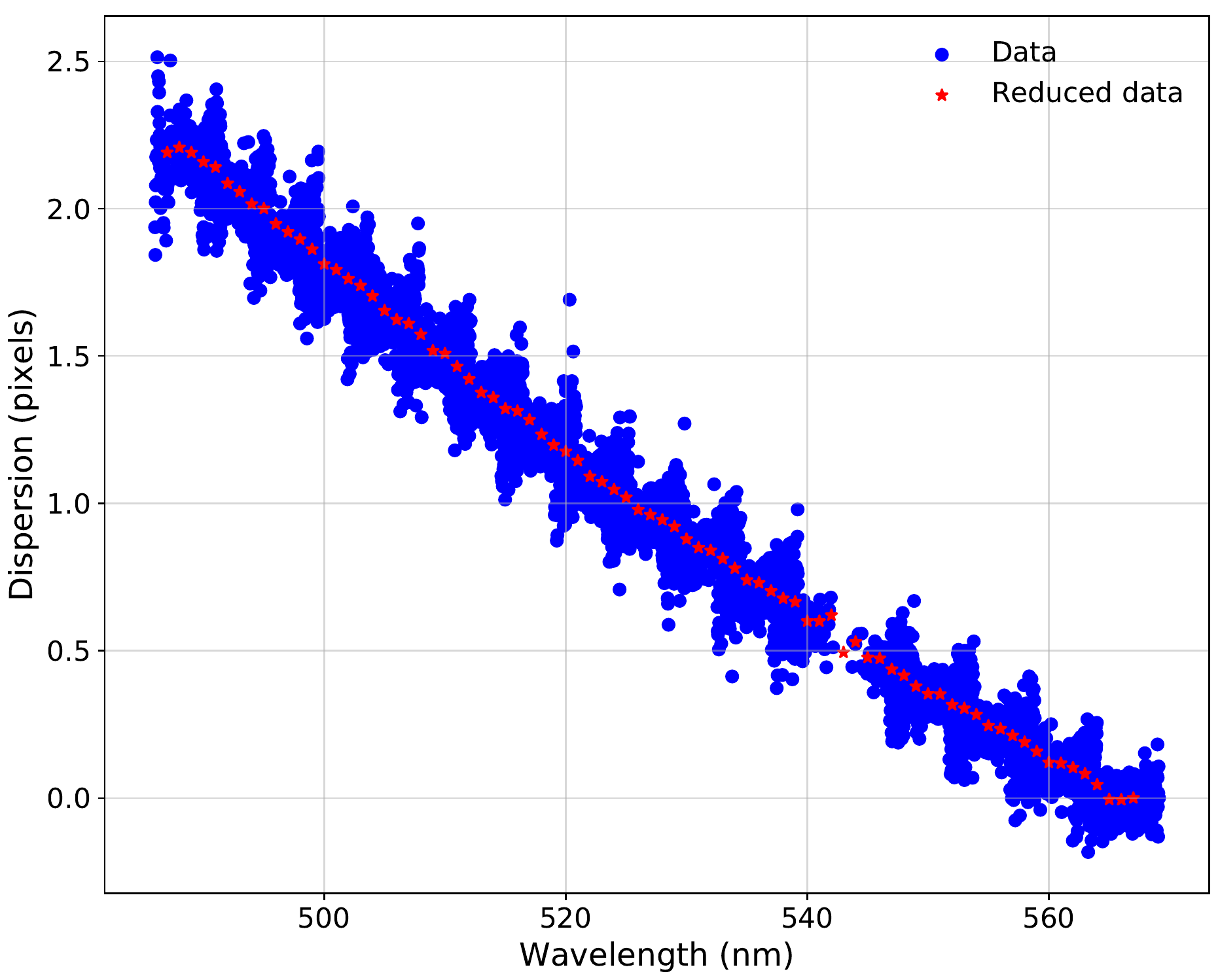}
	\includegraphics[width=\hsize]{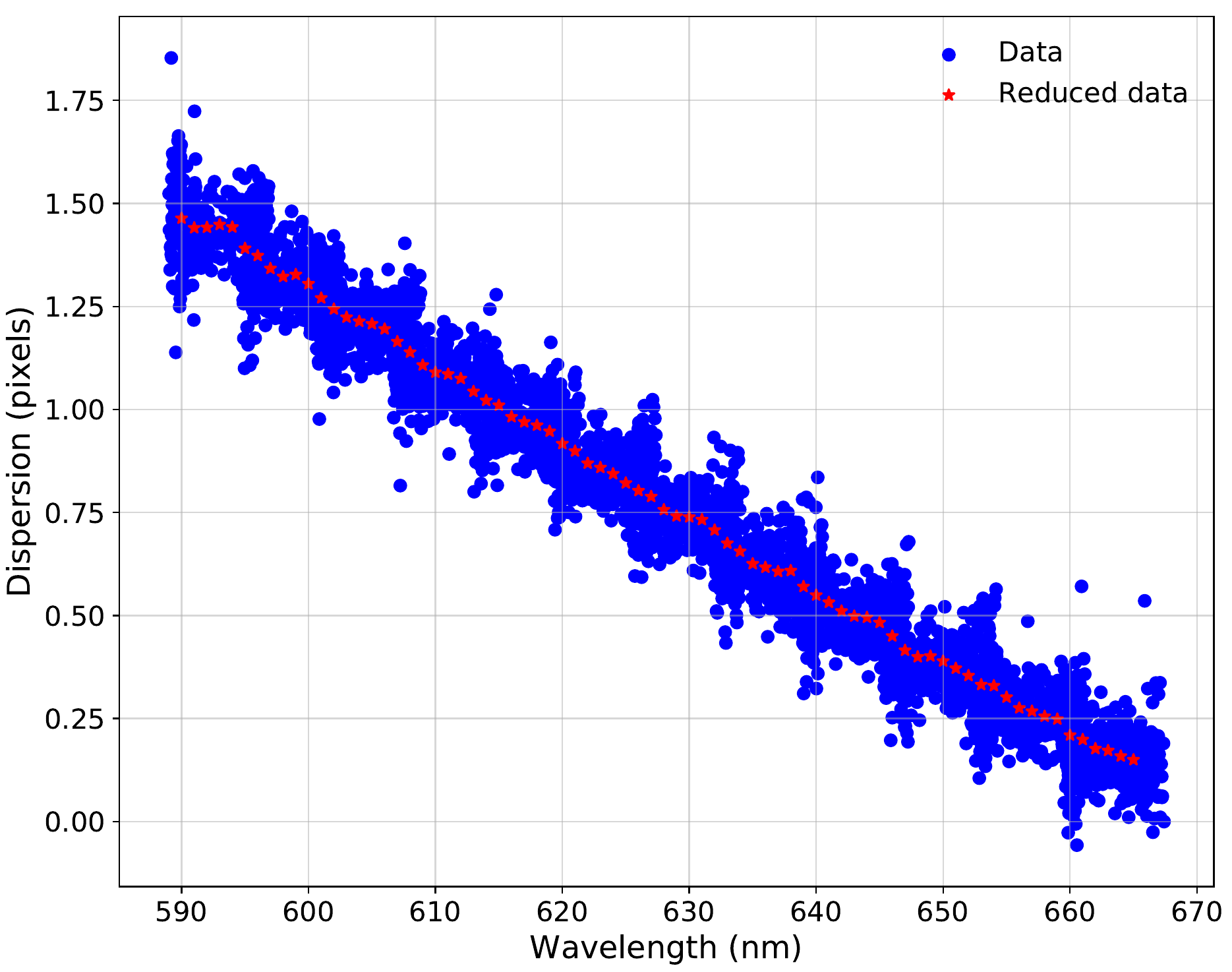}
	\caption{Extracted atmospheric dispersion in pixels using the centroids as a function of wavelength for the red arm. The red stars represents the reduced version of the data after averaging points for each nm. Top: for the range 487 nm to 567 nm; bottom: 590 nm to 665 nm.}
	\label{Fig:notreduced_red}
\end{figure}

\begin{figure}
	\centering
	\includegraphics[width=\hsize]{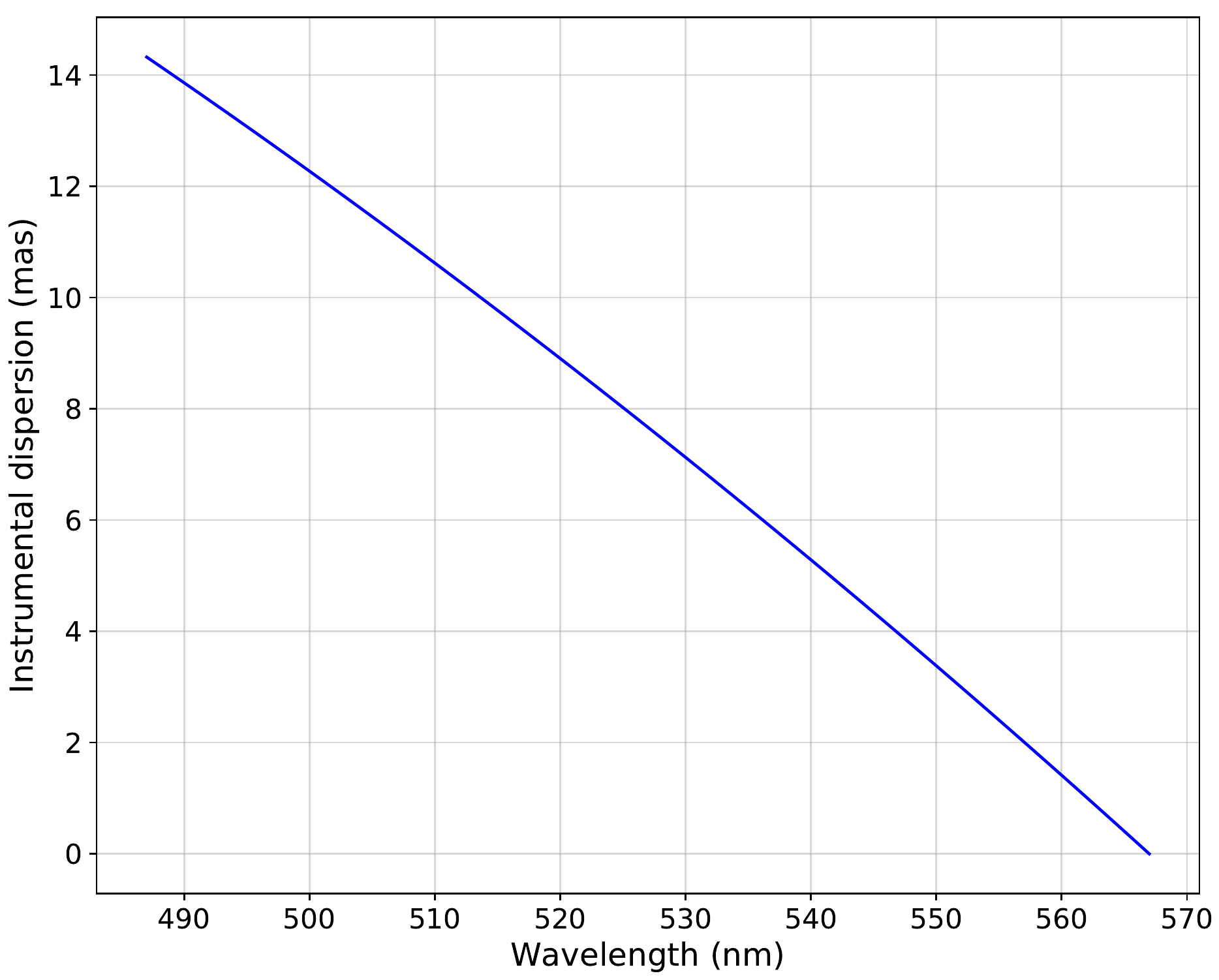}
	\includegraphics[width=\hsize]{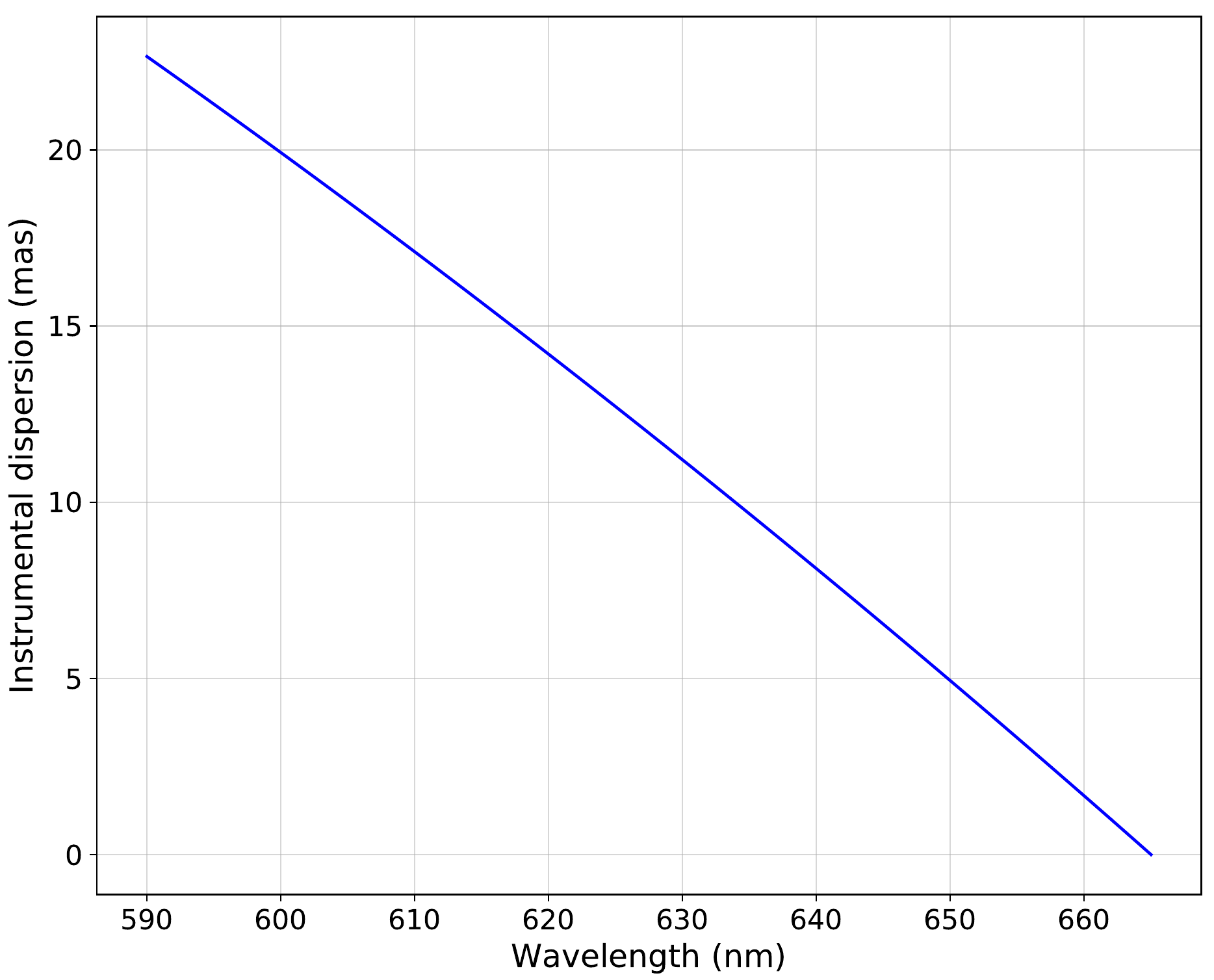}
	\caption{Fit of the instrumental dispersion of UVES using our setup, in the red arm. Top: for the range 487 nm to 567 nm; bottom: 590 nm to 665 nm.}
	\label{Fig:instrumental_red}
\end{figure}

\begin{figure}
	\centering
	\includegraphics[width=\hsize]{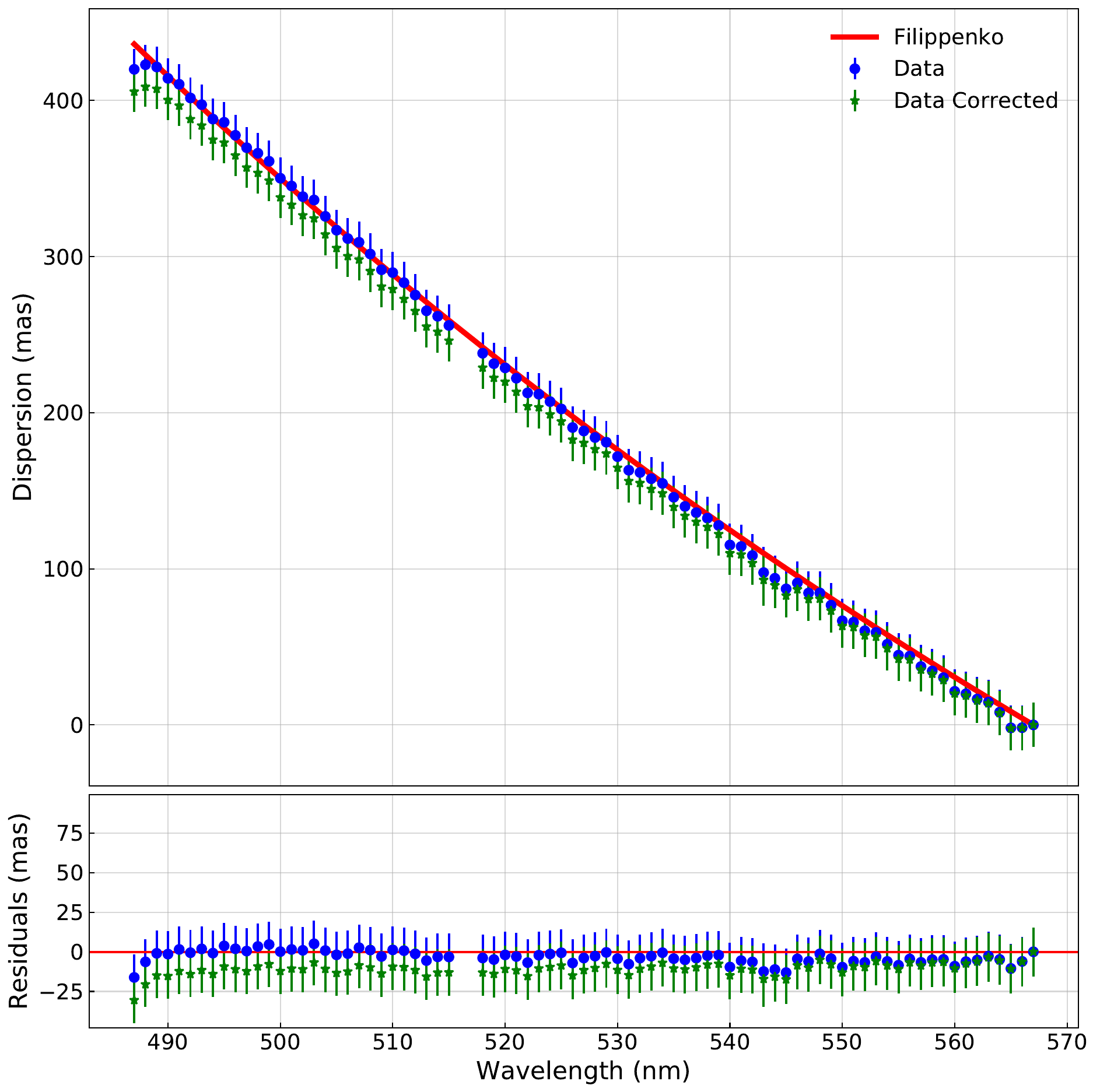}
	\caption{Top: measured atmospheric dispersion, before and after correction by the instrumental dispersion in the red range (487 nm to 567 nm). We also show the expected dispersion using Filippenko's model; bottom: atmospheric dispersion residuals, before and after correction by the instrumental dispersion.}
	\label{Fig:red_all1}
\end{figure}

\begin{figure}
	\centering
	\includegraphics[width=\hsize]{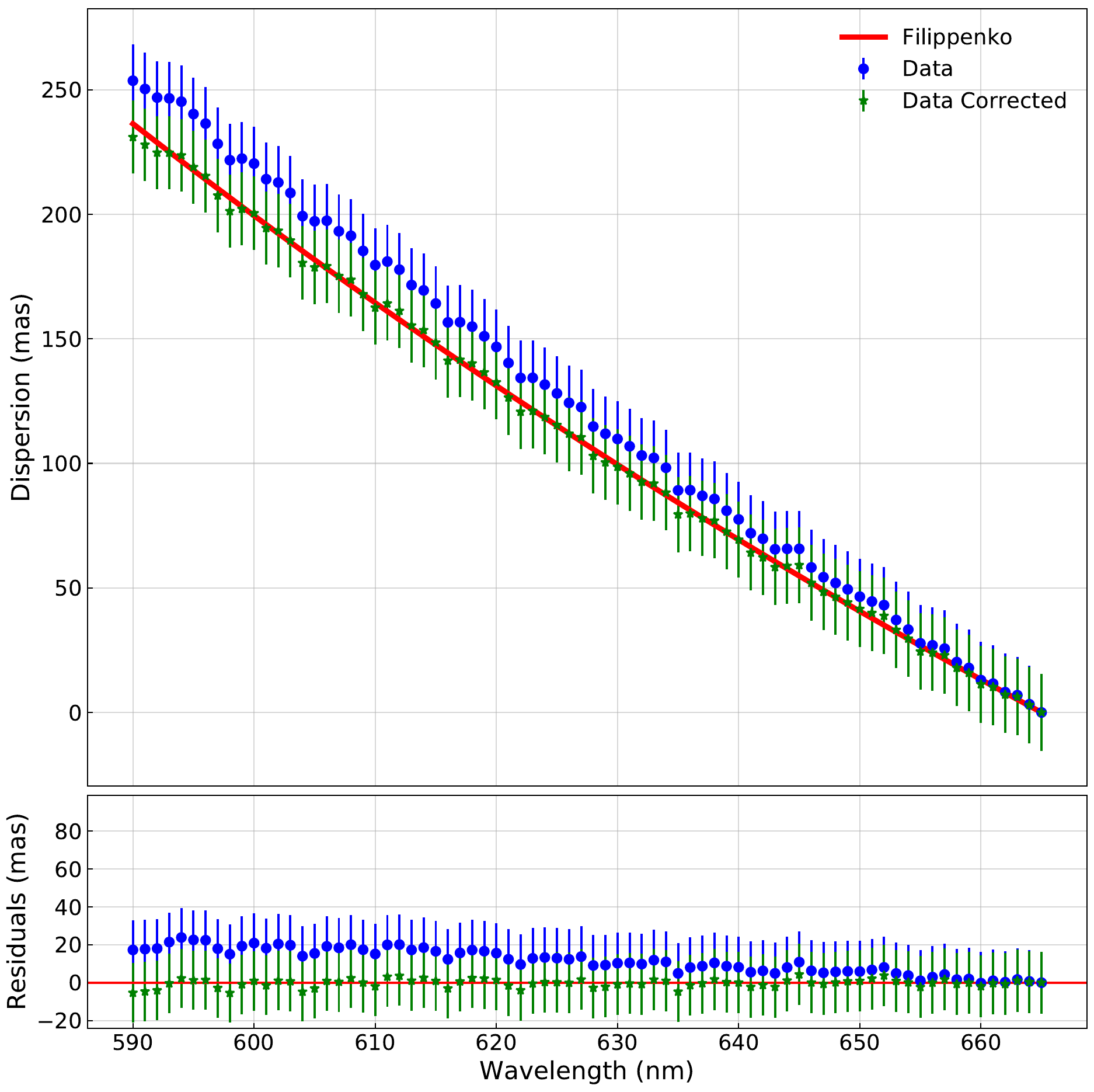}
	\caption{Top: measured atmospheric dispersion, before and after correction by the instrumental dispersion in the red range (590 nm to 665 nm). We also show the expected dispersion using Filippenko's model; bottom: atmospheric dispersion residuals, before and after correction by the instrumental dispersion.}
	\label{Fig:red_all2}
\end{figure}

\section{Angular separation}
\label{appendix:B}
In this appendix we will present the steps to compute the angular separation of a binary-star target.
\begin{enumerate}
	\item Calculate the mean anomaly, M:
	\begin{equation}
		M = \dfrac{360Y}{T} \; degrees,
	\end{equation}
	where $Y$ is the number of years since the epoch of periastron, and $T$ is the period of revolution.

	\item  Solve Kepler's equation:
	\begin{equation}
		E - e \sin{E} = M \; radians,
	\end{equation}
	where $e$ is the eccentricity of the orbit.

	\item  Find the true anomaly, $v$, and the radius vector $r$:
	\begin{equation}
		v = 2\tan^{-1}\left (\sqrt {\left( \dfrac{1+e}{1-e} \right )}\tan \left ( \dfrac{E}{2}\right) \right ),
	\end{equation}
	and
	\begin{equation}
		r = a(1-e\cos{E})
	\end{equation}
	\item Finally:
	\begin{equation}
		\theta = \tan^{-1}  \left\{\dfrac{\sin(v + \omega)\cos{i}}{\cos(\theta + \omega)} \right\} + \Omega ,
	\end{equation}
	and 
	\begin{equation}
		\rho = \dfrac{r \cos(v+\Omega)}{\cos(\theta - \Omega)},
	\end{equation}
	where $i$ is the inclination of the orbit to the plane of the sky, $\omega$ is the longitude of the periastron, and $\Omega$ is the position-angle of the ascending node. $\rho$ is the angular separation of the binary-star target in mas. It is equal to 4.68" for the selected binary used in this work.
\end{enumerate}	 

\bsp	
\label{lastpage}
\end{document}